\documentclass[sigconf]{acmart}

\usepackage{multirow}
\usepackage{tabularx}

\renewcommand\footnotetextcopyrightpermission[1]{} 
\AtBeginDocument{%
  }

\setcopyright{acmlicensed}
\copyrightyear{2018}
\acmYear{2018}
\acmDOI{XXXXXXX.XXXXXXX}

\acmConference[Conference acronym 'XX]{Make sure to enter the correct
  conference title from your rights confirmation emai}{June 03--05,
  2018}{Woodstock, NY}
\acmISBN{978-1-4503-XXXX-X/18/06}

\begin{document}
\title{A Comprehensive Survey on EEG-Based Emotion Recognition:  A Graph-Based Perspective} 

%

\author{Chenyu Liu}
\authornote{Co-author.}
\affiliation{%
  \institution{College of Computing and Data Science, Nanyang Technological University}
  \country{}
}
\email{chenyu003@e.ntu.edu.sg}

\author{Xinliang Zhou}
\authornotemark[1]
\affiliation{%
  \institution{College of Computing and Data Science, Nanyang Technological University}
  \country{}
}
\email{xinliang001@e.ntu.edu.sg}

\author{Yihao Wu}
\affiliation{%
  \institution{College of Computing and Data Science, Nanyang Technological University}
  \country{}
}
\email{yihao005@e.ntu.edu.sg}

\author{Yi Ding}
\affiliation{%
  \institution{College of Computing and Data Science, Nanyang Technological University}
  \country{}
}
\email{ding.yi@ntu.edu.sg}

\author{Liming Zhai}
\affiliation{%
 \institution{School of Computer Science, Central China Normal University}
 \country{}
}
\email{limingzhai@ccnu.edu.cn}

\author{Kun Wang}
\authornote{Corresponding author.}
\affiliation{%
  \institution{University of Science and Technology of China}
  \country{}
}
\email{wk520529@mail.ustc.edu.cn}

\author{Ziyu Jia}
\authornotemark[2]
\affiliation{%
  \institution{Institute of Automation, Chinese Academy of Sciences}
  \country{}
  }
\email{jia.ziyu@outlook.com}

\author{Yang Liu}
\affiliation{%
  \institution{College of Computing and Data Science, Nanyang Technological University}
  \country{}
  }
\email{yangliu@ntu.edu.sg}

\renewcommand{\shortauthors}{Trovato et al.}

\begin{abstract}
  Compared to other modalities, electroencephalogram (EEG) based emotion recognition can intuitively respond to emotional patterns in the human brain and, therefore, has become one of the most focused tasks in affective computing. 
  The nature of emotions is a physiological and psychological state change in response to brain region connectivity, making emotion recognition focus more on the dependency between brain regions instead of specific brain regions.
  A significant trend is the application of graphs to encapsulate such dependency as dynamic functional connections between nodes across temporal and spatial dimensions.
  Concurrently, the neuroscientific underpinnings behind this dependency endow the application of graphs in this field with a distinctive significance.
  However, there is neither a comprehensive review nor a tutorial for constructing emotion-relevant graphs in EEG-based emotion recognition. 
  In this paper, we present a comprehensive survey of these studies, delivering a systematic review of graph-related methods in this field from a methodological perspective.
  We propose a unified framework for graph applications in this field and categorize these methods on this basis.
  Finally, based on previous studies, we also present several open challenges and future directions in this field.

\end{abstract}




\received{20 February 2007}
\received[revised]{12 March 2009}
\received[accepted]{5 June 2009}

\maketitle

\section{Introduction}
Emotion is an integral and complex aspect of human cognition, playing a crucial role in decision-making, behavior, and social interactions \cite{adolphs2003cognitive}. Consequently, emotion recognition is essential for mental health diagnosis and human-computer interaction \cite{zotev2020emotion}. In this context, the direct correlation between electroencephalogram (EEG) signals and brain activity has established EEG-based emotion recognition as a highly specific and valuable task.
First, EEG is a manifestation rather than an expression of emotion, which is more likely to reveal the genuine emotional state than other modalities objectively. The expression of emotion includes facial micro-gestures \cite{toisoul2021estimation}, embodied behavior \cite{de2006towards}, gesture \cite{kita2023gesture}, speech intonation and voice quality \cite{giordano2021representational}, which serve a communicative function and are largely under conscious deliberate control \cite{heaven2020faces}. 
In addition, EEG recordings provide direct measures of neural activity, offering a more accurate representation of brain emotional states than other physiological manifestations of emotion.
Physiological metrics such as skin conductivity, temperature, heart rate, etc., are manifestations of physiological systems in emotional states that do not reflect emotional activity in the human brain directly.
Therefore, EEG-based emotion recognition tasks hold irreplaceable significance in the study of genuine human emotions and the exploration of emotional activities within the human brain.

\begin{figure}[h]
  \centering
  \includegraphics[width=0.9\linewidth]{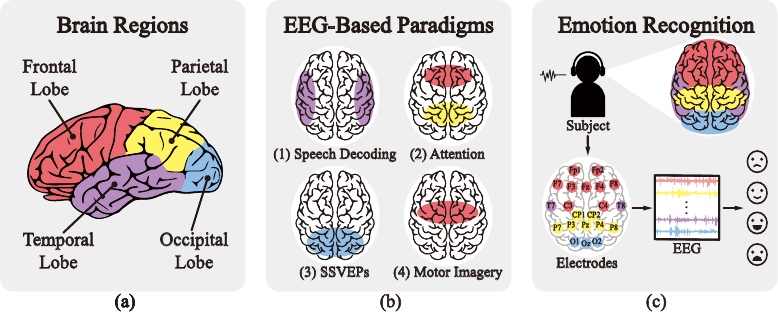}
  \vspace{-2mm}
  \caption{Description of brain regions and EEG-based paradigms. (a) Four regions of the cortex. (b) Four other common EEG-based paradigms. (c) The process of EEG-based emotion recognition.}
  \label{fig:intro}
\end{figure}

In the field of EEG-based research, emotion recognition requires more attention to dependency between brain regions compared to other paradigms \cite{klepl2024graph}. As illustrated in Figure~\ref{fig:intro} (a), the human brain functions as a complex network with hierarchical and functional organization at the level of brain regions \cite{power2011functional}. These regions are responsible for processing specific functions and are particularly relevant to various EEG-based paradigms, as shown in Figure~\ref{fig:intro} (b).
Specifically, the speech decoding paradigm focuses on the temporal lobe, which is responsible for language comprehension and processing \cite{speech_intro}. The primary motor cortex, located in the frontal lobe, regulates motor functions and is thus relevant to motor imagery paradigms \cite{mi_intro}. Visual stimulation activates the occipital lobe, the primary brain region of interest for the steady-state visually evoked potentials (SSVEP) paradigm \cite{ssvep_intro}.
However, as shown in Figure~\ref{fig:intro} (c), emotions are high-level neurocognitive functions expressed as cognitive-emotional interactions generated from brain regions with a high degree of connectivity \cite{pessoa2008relationship}. 
These interconnected brain regions do not function independently \cite{friederici2017language}. Activating a particular brain region often leads to the activation of other regions within the connected network \cite{kober2008functional}.
Thus, developing neurophysiologically meaningful networks to effectively model the connectivity among functional brain regions during emotional states is central to the field.

A significant trend in modeling the aforementioned connectivity is the application of graphs. 
Graphs excel at representing relational data through nodes and edges, making them particularly suitable for modeling the complex interactions between brain regions during emotional states. 
The local or distributed processes between brain regions during emotional states \cite{lettieri2019emotionotopy} can be effectively modeled by edge-based node aggregation \cite{klepl2024graph}, where the edge represents the connectivity between brain regions.
As a result, an increasing number of methods have employed graphs to capture the intricate connectivity patterns inherent in emotion-related EEG, which are crucial for enhancing the accuracy of emotion recognition \cite{grana2023review}. These approaches are specifically developed from diverse perspectives to integrate brain physiology paradigms, distinguishing them structurally from graph-based techniques in other fields. Nonetheless, there is currently no standardized framework for the application of graphs in EEG-based emotion recognition. This lack of standardization highlights the need for a comprehensive survey that encompasses various graph-based methodologies in this field.

To this end, this paper presents a systematic survey of graph applications in EEG-based emotion recognition. Our aim is to offer comprehensive guidance on constructing graph-based models in this field. The contributions of this survey are summarized as follows:
\begin{itemize}
\item \textbf{The first survey.} This survey provides a comprehensive and systematic review of graph-related methods in EEG-based emotion recognition. To the best of our knowledge, this is the first and only survey work on such a topic.
\item \textbf{Systematic methodology-centric taxonomy.} This survey introduces a unified framework from a methodological standpoint on graph application and provides a clear guideline for the construction of graph-related methods in this field.
\item \textbf{Future research directions.} This survey summarises and highlights future directions to facilitate graph application in this field.
\end{itemize}

\begin{figure}[h]
  \centering
  \includegraphics[width=0.8\linewidth]{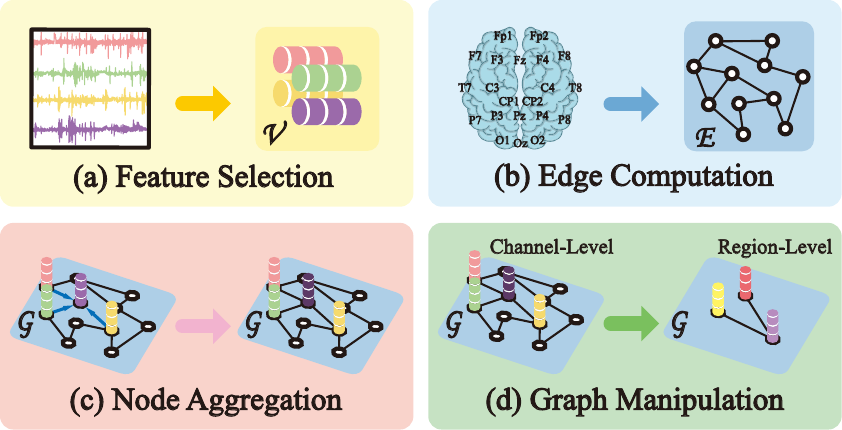}
  \vspace{-2mm}
  \caption{A Unified framework of graph-related methods in EEG-based emotion recognition.}
  \label{fig:stages}
\end{figure}

\section{Background}
\subsection{EEG-based Emotion Recognition}
EEG-based emotion recognition leverages the close relationship between EEG signals and brain activity to accurately identify emotional states. As shown in Figure~\ref{fig:intro} (c), emotions are complex neurocognitive processes involving dynamic interactions among multiple brain regions, such as the frontal lobes, parietal lobes, and temporal lobes \cite{ray2012anatomical}. Each brain region's activity during emotional states contributes uniquely to the overall emotional experience, with specific areas playing distinct roles in the processing of different emotions \cite{herrington2005emotion,salzman2010emotion,bramson2023anxious}. These regions do not operate in isolation; rather, they interact in a highly coordinated manner, forming a network of activity that underlies the processing and regulation of emotions \cite{pessoa2010emotion}. The various channels of an EEG can measure the neural activity of these corresponding regions, capturing real-time changes in brain dynamics \cite{alotaiby2015review}. Thus, emotion-related EEG can reflect the functional connectivity between different brain regions during emotional states, providing a comprehensive view of how emotions are processed in the brain.

\subsection{Preliminaries}
Graph-related methods in EEG-based emotion recognition can be indicated as taking EEG signal $\mathbf{X} \in \mathbb{R}^{C \times S}$ as input, where $C$ is the number of channels and $S$ is the number of samples, constructing graphs and predicting emotion labels $\mathbf{Y} \in \mathbb{R}^{1}$. 
We consider an emotion-related graph denoted as $\mathcal{G} = \{\mathcal{V}, \mathcal{E}\}$, where $\mathcal{V}$ and $\mathcal{E}$ represent the sets of nodes and edges, respectively. 
The feature matrix of $\mathcal{V}$ is denoted as $\mathbf{V} \in \mathbb{R}^{C \times D}$, where $D$ is the dimension of node features. 
An adjacency matrix $\mathbf{A} \in \mathbb{R}^{C \times C}$ represent the connections between nodes, where $e_{i,j} = \mathbf{A}[i,j]$ if $(v_i, v_j) \in \mathcal{E}$. To learn the node representations in a graph $\mathcal{G}$, most methods adhere to the following paradigm of neighborhood aggregation and message passing:

\begin{equation}
\mathbf{h}_i^{(l)} = \text{\fontfamily{lmtt}\selectfont \textbf{COMB}}\left( \mathbf{h}_i^{(l-1)}, \text{\fontfamily{lmtt}\selectfont \textbf{AGGR}}\{  \mathbf{h}_j^{(l-1)}: v_j \in \mathcal{N}(v_i) \} \right),\;0\leq l \leq L
\end{equation}
where $L$ is the number of network layers, $\mathbf{h}_i^{(l)} (1\leq l\leq L)$ denotes the node embedding of $v_i$ at the $l$-th layer. 
{\fontfamily{lmtt}\selectfont \textbf{AGGR}} and {\fontfamily{lmtt}\selectfont \textbf{COMB}} represent functions used for aggregating information from neighborhood nodes $\mathcal{N}(v_i)$ and combining ego- and neighbor-representations, respectively.

\begin{figure*}[htbp]
\centering
\includegraphics[width=2\columnwidth]{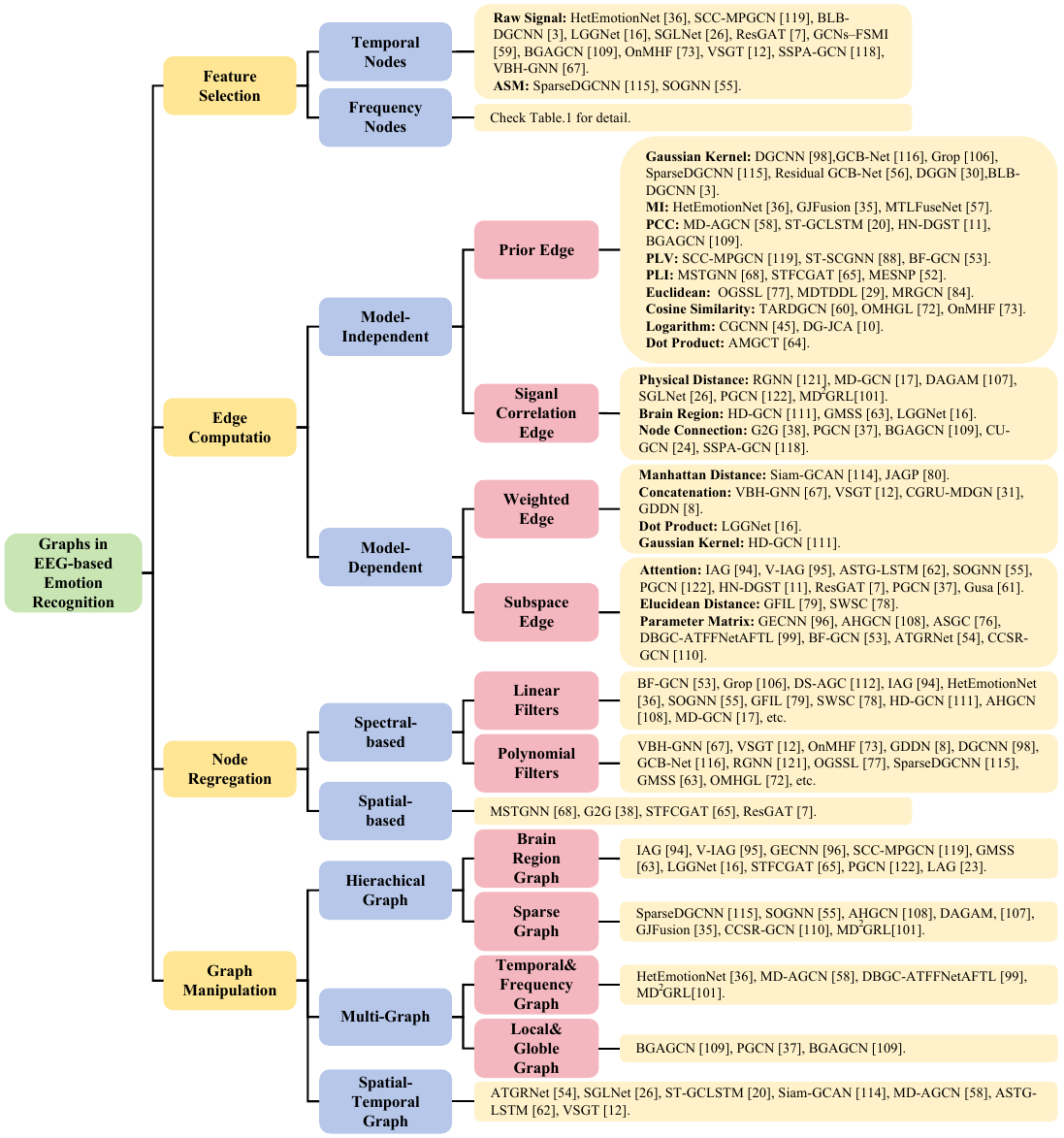}
\vspace{-3mm}
\caption{An overview of the categorization.}
\centering
\label{fig:class}
\end{figure*}

\section{Taxonomy}
Recent advancements in graph application have demonstrated their great potential to achieve better performance in EEG-based emotion recognition.
These methods focus on different aspects of the graph but generally address a few key questions: (1) what kind of EEG features should be selected; (2) how to compute the adjacency matrix to characterize the connectivity of brain regions in emotional states; (3) how to perform node aggregation; and (4) what kind of graph manipulation should be adopted. 
Based on these questions, we summarized and categorized the existing studies based on a unified framework as shown in Figure~\ref{fig:stages}.

The proposed taxonomy is illustrated in Figure~\ref{fig:class}, and the related works can be found in Table~ \ref{tab:baseline}. 
The proposed taxonomy offers a structured and comprehensive classification to deepen the understanding of graph-based approaches in EEG-based emotion recognition. It is organized into four hierarchical levels, starting with the feature selection, followed by the edge computation, node aggregation, and finally, the graph manipulation. 
\textbf{(a) Feature selection} indicates the type of node features selected, which contains temporal and frequency nodes;
\textbf{(b) Edge computation} specifically refers to calculating the adjacency matrix, which is further classified into model-dependent and model-independent edges according to the participation of parameters.
\textbf{(c) Node aggregation} denotes the updating method of nodes, which is categorized into spectral-based and spatial-based methods.
\textbf{(d) Graph manipulation} represents the manipulation of graph structures at the network level, including multi-graphs, hierarchical graphs and spatio-temporal graphs.
In the following, we introduce the four stages and their subcategories in more detail.

\section{Feature Selection}
\label{sec: node}
The unique characteristics of emotion-related EEG have led to the development of distinct methods for node feature selection in graph-based approaches within this field. As EEG is a time series data, the most intuitive approach is to utilize temporal features. 
Additionally, due to the direct correlation between EEG frequency and emotional activity \cite{bijanzadeh2022decoding}, some methods employ unique approaches to extract frequency domain features as nodes.
In this section, we classify the types of nodes employed in graph-based methods into temporal and frequency nodes. In the following, we provide a detailed introduction to these two types.

\textbf{Temporal Nodes:} This is the most intuitive and straightforward method, where nodes directly use raw signals or time-related features. For example, LGGNet \cite{LGGNet} and SCC-MPGCN \cite{SCC-MPGCN} input the raw EEG signals into the network after basic filtering. 
Raw EEG signals facilitate joint analysis with other physiological signals. HetEmotionNet \cite{HetEmotionNet} and VBH-GNN \cite{VBH-GNN} incorporate temporal physiological signals such as Electrocardiogram (ECG) and Galvanic Skin Response (GSR) as auxiliary modalities for synchronous time-domain analysis. Additionally, Sparse-DGCNN \cite{Sparse-DGCNN} employs Amplitude Spectrum Mean (ASM), using the average amplitude spectrum of EEG signals as node features.

\textbf{Frequency Nodes:} The fundamental reason for utilizing frequency features as nodes is that there exists a direct connection between different frequency bands of EEG and different emotions \cite{mcclay2023dynamic, bijanzadeh2022decoding}, such as the $\beta$-band is associated with anxiety. 
Therefore, frequency nodes typically focus on five specific bands: delta (1-3 Hz), theta (4-7 Hz), alpha (8-13 Hz), beta (14-30 Hz), and gamma (31-50 Hz) bands.
The most commonly used frequency node at present is the Differential Entropy (DE) feature \cite{duan2013differential}:


\begin{equation}
\begin{gathered}
DE(\mathbf{X})  =-\int_{-\infty}^{\infty} \frac{1}{\sqrt{2 \pi \sigma^2}} e^{-\frac{(x-\mu)^2}{2 \sigma^2}} \log \left(\frac{1}{\sqrt{2 \pi \sigma^2}} e^{-\frac{(x-\mu)^2}{2 \sigma^2}}\right) d x \\
 =\frac{1}{2} \log \left(2 \pi e \sigma^2\right),
\end{gathered}
\end{equation}

where $\mu$ and $\sigma$ denote the parameters of Gaussian distribution $\mathcal{N}\left(\mu, \sigma^2\right)$ that the EEG obeys. Many methods, such as \cite{MD-GCN,STFCGAT,OGSSL}, estimate the probability density function of the signal and use the aforementioned formula to calculate the DE for each band as node features.
Furthermore, methods such as \cite{V-IAG,Gusa,GCBNet} utilize combinations of DE features from symmetrical electrodes (differential asymmetry and rational asymmetry).
In addition, methods such as \cite{OMHGL,G2G,ASTG-LSTM} use the Power Spectral Density (PSD) of EEG signals. A few methods employ other frequency domain features, such as the Differential Cumulative Average of Uniformity (DCAU) used by DGCNN \cite{DGCNN} and the Short-Time Fourier Transform (STFT) used by GDDN \cite{GDDN}. \\ 

\noindent \textit{Remark.} Although the majority of current research employs frequency nodes, we argue this prevalence is not due to the superiority of frequency nodes over temporal nodes but is instead influenced by the datasets. For instance, methods using the SEED \cite{seed} dataset commonly adopt DE features, while those using the MPED \cite{mped} dataset typically use STFT. Only a few methods utilizing the DREAMER \cite{dreamer} and DEAP \cite{deap} datasets employ temporal nodes. 
The high usage rate of the SEED dataset has contributed to the prevalence of frequency nodes.
It is evident that frequency nodes can mitigate the effect of transient noise (e.g., eye movements) across multiple frequency bands, thereby providing a more stable representation of emotional states. However, apart from their end-to-end implementation, temporal nodes exhibit greater potential in reflecting event-related potentials in EEG under emotional activity, such as transient changes in amplitude and waveform \cite{schupp2003attention}. Additionally, temporal nodes offer higher extensibility, as they can accommodate other physiological signals for joint analysis \cite{wang2024research}.

\section{Edge Computation}
\label{sec: edge}
Edge computation poses a significant challenge for graph-based methods in EEG-based emotion recognition. On the one hand, the edges in emotion-related graphs have inherent neurophysiological foundations. For instance, the electrode positions of EEG acquisition devices generally conform to the 10-20 system, which dictates fixed distances and connectivity between nodes, each representing different brain regions. On the other hand, the dependencies between brain regions under emotional states are highly variable. 
Firstly, there are differences in brain region dependencies across different emotional states \cite{pessoa2008relationship}. 
Secondly, even under the same emotional state, different subjects exhibit variations in the intensity and patterns of their emotional brain activities \cite{lindquist2012brain}.
Additionally, for the same subject, brain activity patterns can vary when experiencing the same emotion at different times due to differences in response intensity and psychophysiological states \cite{tambini2017emotional}.
Therefore, graph-based methods in this field strive to balance the fixed and variable nature of edges. In this section, we categorize these methods into two types based on whether model parameters are involved in edge computation: model-independent and model-dependent methods. In the following, we describe these two methods and their subtypes in detail.

\subsection{Model-Independent Edge}
The computation of model-independent edges does not involve model parameters. These edges aim to maximally preserve the neurophysiological foundations of emotion-related edges, thus having clear physiological structures and physical significance. The existing literature typically adopts the following two strategies.

\textbf{Prior Edge:} In this method, edges are manually defined based on electrode connections from physiological paradigms. 
An example of prior edges is in \cite{RGNN, DAGAM, MD-GCN}, which utilize the theory that the strength of connections between brain regions decays as a function of the inverse square of the physical distance \cite{salvador2005neurophysiological}. These methods use 3-D coordinates to calculate the physical distances between electrodes as the edges. Methods such as \cite{G2G, BGAGCN, PGCN} determine the connectivity between electrodes directly based on the electrode arrangement in the International 10-20 system, thereby constructing the adjacency matrix. Based on this, \cite{LGGNet, HD-GCN, GMSS} restrict the connectivity between electrodes within specific groups, considering the distribution of brain regions and the symmetry between the left and right hemispheres.

\textbf{Siganl Correlation Edge:} The computation in this method is based on signal processing, using the similarity or distance between EEG signals from different channels as edges.
The propagation of physiological electrical signals is diffusive \cite{pang2023geometric}, which means that the signal collected by one electrode will contain EEG signals from adjacent electrodes.
Therefore, signal correlation edges preserve the neurophysiological foundations between electrodes through signal-level similarity.
\cite{OGSSL, MDTDDL, MRGCN} use the Euclidean distance between node signals as edges. Similarly, \cite{TARDGCN, OMHGL} employ cosine distance, while MSTGNN \cite{MSTGNN} uses the Phase Lag Index (PLI). Additionally, \cite{ST-GCLSTM, MD-AGCN, HN-DGST, BGAGCN} utilize the Pearson Correlation Coefficient (PCC) to compute the strength of the linear relationship between signals as edges. \cite{SCC-MPGCN, MESNP, BF-GCN, ST-SCGNN} use the Phase Locking Value (PLV) to calculate the degree of phase similarity between electrode signals at a particular frequency as their edges. Other methods include Gaussian kernel, such as 
\cite{Sparse-DGCNN, GCBNet, DGCNN}, and Mutual Information (MI), as used by \cite{HetEmotionNet, GJFusion, MTLFuseNet}. \\ 

\noindent \textit{Remark.} The limitation of model-independent edges is that, although they preserve the basic connectivity of electrodes under emotional states, it is challenging to model the synergistic relationships between brain regions during emotional activities. Even the relatively flexible signal correlation edges, which use data-driven computation methods to reflect differences between different individuals and emotional states, are still insufficient to capture the complex spatial-temporal dependencies. However, the relatively fixed learning approach of model-independent edges does not mean they cannot be optimized during training. A common update method is:
\begin{align}
\mathbf{A'}=(1-\rho) \mathbf{A}+\rho \frac{\partial \text { Loss }}{\partial \mathbf{A}},
\end{align}
where $\rho$ denotes the learning rate. This allows model-independent edges to adapt to the variability of EEG under emotional states to some extent and ensures stable performance in scenarios with limited training data.

\subsection{Model-Dependent Edge}
The introduction of parameters allows model-dependent edges to dynamically fit the complex dependencies in emotion-related EEG. Model-dependent edges can adjust and refine the relationships between nodes, which represent different brain regions, based on the varying intensities and patterns of emotional activities. The existing literature typically adopts the following two strategies.

\textbf{Weighted Edge:} In this method, model parameters are introduced as a weight matrix in the computation, transforming the connection between nodes into nonlinear ones through activation functions. It can be regarded as a special case of signal correlation edges combined with model parameters. A common computation method for weighted edges is:
\begin{align}
e_{ij}=\frac{\exp \left(\sigma\left(\omega^T\mathcal{R}\{xi,xj\}\right)\right)}{\sum_{j=1}^C \exp \left(\sigma\left(\omega^T\mathcal{R}\{xi,xj\}\right)\right)}
\end{align}
where $\sigma(\cdot)$ represents an activation function, such as $\operatorname{ReLU}$. $x_i$ and $x_j$ represent the node embeddings of $v_i$ and $v_j$, and $\omega$ denotes the model parameters. $\mathcal{R}\{\cdot,\cdot\}$ represents a specific operation on the node embedding pairs. For example, in \cite{Siam-GCAN, JAGP}, it represents the computation of the Manhattan distance, while in \cite{MSFR-GCN, DS-AGC}, it represents the computation of the Euclidean distance. Additionally, some methods, such as \cite{GDDN, CGRU-MDGN}, concatenate the node embeddings directly for linear transformation, with VBH-GNN concatenating the node embeddings along the channel dimension and using convolutional layers instead of linear transformations to reduce the number of parameters.

\textbf{Subspace Edge:} This method uses model parameters to project the node embeddings into a subspace, where the dot product of the projected nodes is used as edges. The process can be described as:
\begin{align}
e_{i,j}=\sigma\left((x_i \omega_{i})(x_j \omega_{j})^T\right),
\end{align}
where $\omega_{i}$ and $\omega_{j}$ represent the projection matrices for two nodes, respectively. For example, \cite{PGCN,V-IAG,IAG} adopt the above computation formula. Fundamentally, the dot product in the projection space is consistent with the dot product of the Query and Key matrices in the attention mechanism. Therefore, some methods, such as \cite{ASTG-LSTM, HN-DGST, SOGNN, ResGAT}, directly use the output of the attention mechanism as edges. \\

\noindent \textit{Remark.} Model parameters allow these edges to dynamically fit the complex spatial-temporal dependencies in emotion-related EEG.
For example, as a special case of signal correlation edges, weighted edges create more flexible nonlinear connections. However, the introduction of more model parameters increases the risk of overfitting. 
This is particularly true for subspace edges, which lack specific node relationship constraints, making it difficult to generalize in scenarios with insufficient training data. 
Additionally, some methods, such as \cite{ASGC, ATGRNet, CCSR-GCN, AHGCN}, randomly initialize a parameter matrix as the adjacency matrix. This entirely data-independent approach is even more challenging to train.
Although many previous studies have not opted for model-dependent edges in subject-dependent scenarios, we argue that this parameterized, data-driven edge computation method is the future trend for EEG-based emotion recognition. For example, for cross-dataset or cross-subject domain adaptation scenarios or emotion foundation models, adopting model-dependent edges can more accurately capture the underlying dependencies in emotion-related EEG data. Moreover, one potential approach is to integrate more neurophysiological foundations into model-dependent edges. For instance, VSGT \cite{VSGT} simulates the superposition state of neuronal electrical signals using Gaussian approximation and introduces parameters through re-parameterization \cite{kingma2013auto} to update model-dependent edges.

\section{Node Aggregation}
\label{sec: agg}
Graph-based methods in this field commonly utilize convolutional operations to aggregate node features. These methods generate representations for node $v_i$ by aggregating its own features $h_i$ and the features of its neighbors $h_j$ where $v_j \in \mathcal{N}(v_i)$. 
Consequently, these methods employ a fixed number of convolutional graph layers to extract high-level node representations, thereby capturing the dependencies between brain regions under emotional states. 
We categorize these methods into spectral-based and spatial-based approaches. In the following sections, we delve into the specifics of these two categories and their subtypes.

\subsection{Spectral-based Method}
Spectral-based methods capture the dependencies between brain regions under emotional states by transforming all node features into a weighted sum of different feature vectors through the graph Fourier transform. This means that the features of all EEG channels are filtered through a fully connected feature space, with messages being simultaneously transmitted between all nodes, thereby simulating the global coordination between brain regions under emotional states. We define the convolution process of this aggregation method as:
\begin{align}
h_i^{(l)}=\sigma\left(\mathbf{U} \sum_{j=1}^{C} \mathbf{G}_{i j}^{(l-1)} \mathbf{U}^{\top} h_j^{(l-1)}\right),
\end{align}
where $\sigma$ is the activation function, $h_i^{(l-1)}$ denotes the node embedding of $v_i$ at the $l$-th layer, and $\mathbf{G}_{i j}^{(l-1)}$ represents a diagonal matrix with learnable parameters.
This process can also be seen as using $\mathbf{G}$ as a filter to remove noise from the emotion-related EEG graph signals. The existing research employs the following two types of filters.

\textbf{Polynomial Filters:} In this method, the filter is replaced by a simplified polynomial function, avoiding explicit eigenvalue decomposition and thereby reducing the time complexity. In this field, all methods using polynomial filters, such as \cite{V-IAG, DGCNN, ATGRNet}, employ Chebyshev expansion mentioned in ChebNet \cite{chebnet} to construct the filter:
\begin{align}
h_i^{(l)}= \sigma(\mathbf{U}(\sum_{k=1}^K \theta_k \mathcal{T}_k(\tilde{\Lambda}))\mathbf{U}^{\top}h_j^{(l-1)})
\end{align}
where $\tilde{\boldsymbol{\Lambda}}=\frac{2 \boldsymbol{\Lambda}}{\lambda_{\max }}-\mathbf{I}_N$ is the eigenvalue matrix scaled in the range $\left[-1,1\right]$. $\boldsymbol{\Lambda}$ is a diagonal matrix of eigenvalues. $\mathcal{T}_k(x)=2 x \mathcal{T}_{k-1}(x)-\mathcal{T}_{k-2}(x)$ constructs the orthogonal space, and $\mathcal{T}_k(\tilde{\Lambda})$ is the $k$-th order Chebyshev polynomial at $\tilde{\boldsymbol{\Lambda}}$.

\textbf{Linear Filters:} Linear filters are a further simplification of polynomial filters. Methods that use linear filters, such as \cite{JAGP, DS-AGC, ST-SCGNN}, similar to GCN \cite{GCN}, adopt the first-order approximation of the Chebyshev expansion as the filter. Their convolution process can be summarized as follows:
\begin{align}
h_i^{(l)}=\sigma(\overline{\mathbf{A}} h^{(l-1)}\omega_i),
\end{align}
where $\omega_i$ represents the weight matrix. $\overline{\mathbf{A}}= \mathbf{I}_{N}+\mathbf{D}^{-(1 / 2)} \mathbf{A} \mathbf{D}^{-(1 / 2)}$ is the renormalization trick of the adjacency matrix. Obviously, the graph filter is linear with the input adjacency matrix. From a spatial-based perspective, these methods can be considered as aggregating feature information from a node’s neighborhood. \\

\noindent \textit{Remark.} Currently, there has been limited innovation in the design of filters within graph-based methods in this field. Instead, researchers continue to use established methods such as ChebNet and GCN. However, even GCN, an improved form of ChebNet, has seen numerous advancements and modifications within the graph neural network domain due to its inherent limitations. The brain region activities under emotional states are complex and dynamic, making it challenging for the simple linear filters of GCN to effectively capture the relationships between EEG channels. This is because the predefined filters in GCN are typically designed for first-order information, which cannot directly capture such high-order relationships.
In summary, there is a need for further exploration and improvement in the selection and design of filters, as well as in the development of methods specifically tailored for emotion-related EEG. 
Combining linear filters with complex filters to learn high-order brain region dependencies should be a key focus for EEG-based emotion recognition.

\subsection{Spatial-based Methdos}
Spatial-based methods define node aggregation based on the spatial relationships between nodes. Specifically, these methods convolve the node representation of the central channel with those of its neighboring channels to derive an updated representation for the central channel, thereby propagating information along the connectivity of the electrodes.
The primary distinction between spatial-based and spectral-based methods lies in their respective modes of information propagation between nodes. Methods such as \cite{G2G, STFCGAT, ResGAT} incorporate attention mechanisms during the propagation process to assign weights to the contributions of neighboring nodes to the central node. Consistent with the GAT \cite{GAT}, their update process can be summarized as:
\begin{align}
h_i^{(l)}=\sigma\left(\sum_{v_j \in \mathcal{N}(v_i)} \alpha_{i j} \omega_i h_j^{(l-1)}\right),
\end{align}
where $\alpha_{i j} = \sigma(\text{\fontfamily{lmtt}\selectfont \textbf{ATT}}(h_i^{(l-1)},h_j^{(l-1)}))$ represents the attention weight, or connective strength, between $v_i$ and $v_j$. Similar to linear filters, these methods update node embeddings through linear transformations. However, their adjacency matrices, which denote the existence of edges between nodes, limit the update scope of the central node to its connected neighbors. This results in sparse adjacency matrices that emphasize local spatial dependencies.

Another method is grounded in the theory that there exist inherent connections and pathways in the brain during emotional activities \cite{bullmore2009complex}. 
MSTGNN \cite{MSTGNN} utilizes the minimum spanning tree to simulate pathways between electrodes. 
The pathways are defined as the routes through which electrical signals propagate during emotional activities. Within a pathway, each node can have up to one parent node and multiple child nodes. Consequently, the update method is:
\begin{align}
h_i^{(l)}=\sigma(\omega_i\left(\left(1+\varepsilon^{(l-1)}\right) \cdot h_i^{(l-1)}+\omega_f h_f^{(l-1)}+\right.
\left.\omega_c \text{\fontfamily{lmtt}\selectfont \textbf{MEAN}}\left(h_c^{(l-1)}\right)\right),
\end{align}
where $h_f^{(l-1)}$ represents the node embedding of the parent node. If $v_i$ is the root node, the parent node embedding is set to zero. $h_c^{(l-1)}$ represents the node embeddings of the child nodes $v_c \in \mathcal{N}(v_i)$. \text{\fontfamily{lmtt}\selectfont \textbf{MEAN}} denotes averaging over all child nodes.
$\omega_f$ and $\omega_c$ are learnable parameters for parent and child nodes. \\

\noindent \textit{Remark.} In EEG-based emotion recognition, the boundaries between spatial-based and spectral-based methods are becoming increasingly blurred. This is because node aggregation in spatial-based methods is essentially an approximate linear process, which aligns with the principles of linear filters used in spectral-based methods. Consequently, employing an attention mechanism to infer emotion-related EEG edges followed by node aggregation using spectral-based methods is fundamentally and practically similar to first determining electrode connectivity and then incorporating attention mechanisms within spatial-based aggregation.

A promising approach is to integrate neurophysiological foundations into spectral-based methods. For example, MSTGNN uses Minimum Spanning Trees (MST) to simulate brain pathways under emotional states. This aligns with the perspectives introduced in edge computation, emphasizing the incorporation of neurophysiological foundations into model parameters. 
By doing so, models can more accurately capture the intricate dependencies and interactions that characterize emotional brain activity, enhancing both the robustness and interpretability of EEG-based emotion recognition systems.
In summary, the convergence of spatial-based and spectral-based methods, along with the integration of neurophysiological insights, holds significant potential for advancing the effectiveness of EEG-based emotion recognition models.

\section{Graph Manipulation}
\label{sec: graph}
In EEG-based emotion recognition, graph-related methods frequently implement specific model-level graph manipulations to enhance the model's ability to represent and learn the complex dependencies inherent in the data. These manipulations address the multifaceted nature of emotion-related EEG, which includes variations across both the frequency and temporal domains, as well as intra- and inter-regional brain connections. 
Such structural adjustments are crucial for tailoring graph representations to better align with the neurophysiological foundations of emotional brain activity.
Existing literature frequently adopts the following three strategies:

\subsection{Hierachical Graph}
Hierarchical Graphs leverage the natural grouping of nodes to align with the physiological foundations of emotion-related EEG, particularly the distribution of brain regions. 
This approach aims to construct the spatial dependencies of emotion-related EEG from a more macroscopic perspective. Based on the different types of dependencies established by this method, we categorize existing research into two types.

\textbf{Brain Region Graph:} In this method, channel-level nodes are grouped to further infer inter-regional dependencies at the brain region level. These dependencies reflect the coordination between different brain areas, such as the interactions between the frontal cortex and the amygdala during emotion regulation \cite{salzman2010emotion}. 
One type of method employs a manually established hierarchy based on the distribution of electrodes corresponding to different brain regions. For example, \cite{IAG, V-IAG} infer fully connected spatial dependencies in emotion-related EEG and then divide the brain regions into 17 groups according to the 10-20 system, averaging the embeddings of channels within each group to obtain regional embeddings. Another strategy, adopted by methods such as \cite{GMSS, STFCGAT, GECNN, LAG}, is to first group nodes based on brain region distribution to learn intra-regional spatial dependencies and then use the regional graph embeddings as nodes in a global graph to learn inter-regional relationships. In addition to grouping based on brain regions, some methods like \cite{LGGNet, PGCN} also divide the hierarchy according to the left and right hemispheres of the brain.
The rest method involves adaptively grouping channels and then inferring spatial dependencies between these groups. For instance, \cite{AHGCN, SCC-MPGCN} set a parameter weight matrix to linearly transform the channel-level node embeddings, reducing the dimensionality of the graph embeddings from the number of channels to the number of brain regions to achieve hierarchical division.

\textbf{Sparse Graph:} This method aims to represent the small-world topology \cite{pandit2013traumatic} of brain regions during emotional activities, characterized by densely clustered local connections with sparse long-range connections \cite{fornito2015connectomics}. The underlying theory is that certain brain functions activate only a limited number of brain regions. Thus, sparse cortical activity can explain the EEG patterns generated by deeper sources \cite{krishnaswamy2017sparsity}.
For example, MD$^2$GRL \cite{MD$^2$GRL} uses a parameter as a threshold to filter and retain a specific number of nodes, thereby achieving sparse connectivity among brain regions. Similarly, methods in \cite{DAGAM, SCC-MPGCN} learn a variable weight for all nodes, selecting a specific number of nodes based on ranking. Additionally, conventional pooling methods can achieve similar results, as demonstrated by GJFusion \cite{GJFusion}. \\

\noindent \textit{Remark.} These two types of methods represent efforts in the field to combine neurological foundations with network structures. We argue that integrating them could potentially offer significant benefits. Currently, the approach of retaining a specific number of nodes in sparse graphs inevitably leads to the loss of potentially valuable information. Therefore, basing the model on the brain region graph and then inferring sparsity on the coarsened graph could maximize the retention of channel information while simulating the sparse connectivity of brain regions.

\subsection{Multi-Graph}
Multi-Graph allows models to capture different emotion-related EEG dependencies simultaneously by concatenating multiple types of graph embeddings. This method leverages the complementary information provided by different graph structures, enriching the feature space and improving the model's ability to discern emotional states. 
Based on the complementary domains, we further introduce two types of this method in the following.

\textbf{Temporal\&Frequency Graph:} In this method, the model employs parallel graph structures to build the dependencies of emotion-related EEG in both the temporal and frequency domains. Temporal dependencies enable the model to detect amplitude features that are closely correlated with brain region, while frequency dependencies capture the activation of specific frequency bands associated with different emotional states, such as increased energy in the alpha band during sadness \cite{fan2024brain,shim2018altered}. Methods in  \cite{HetEmotionNet,MD$^2$GRL} integrate temporal\&frequency graphs at the graph embedding level. They utilize a two-stream structure, where each stream corresponds to the temporal-spatial and frequency-spatial domains, respectively, and compute its own adjacency matrix. The graph embeddings from both streams are then concatenated and fed into a classifier. A distinct strategy employed by \cite{MD-AGCN,DBGC-ATFFNetAFTL} involves merging the adjacency matrices of the two streams before the node aggregation, resulting in a shared hybrid adjacency matrix. 

\textbf{Local\&Globle Graph:} This graph is less commonly applied in this field. Since it distinguishes itself from brain region graph methods only by decoupling the inference of inter-regional and intra-regional dependencies into two parallel processes. 
In the Local stream, electrodes are grouped based on brain regions, and connectivity is confined within these regions. Methods such as \cite{HD-GCN,BGAGCN,PGCN2} utilize this strategy to focus on the local dependencies within specific brain areas. MRGCN \cite{MRGCN} replaces the fixed brain region divisions with a more flexible framework. It introduces short-range and long-range spatial dependencies, which correspond to localized intra-region correlations and inter-region correlations, respectively. \\

\noindent \textit{Remark.} Currently, multi-graph methods in the field struggle to ensure that the accuracy improvements gained from multi-stream structures justify the additional computational burden. This challenge arises because most multi-graph implementations simply merge two types of graph embeddings without fully optimizing the interaction between them. However, it is undeniable that multi-graph methods effectively leverage the complementary nature of different dependencies in emotion-related EEG data.

\subsection{Spatial–Temporal graph}
\label{sec:stg}
In this method, emotion-related EEG is decomposed into multiple time slices to construct temporal dependencies between these slices. This method views emotions as dynamic processes rather than static states. By using a temporal encoder, it captures the changes in these spatial dependencies over time to infer emotion labels. 
The difference between the current methods lies mainly in the selection of the encoder.
For example, \cite{ST-GCLSTM, ASTG-LSTM, SGLNet} employ LSTM as the encoder for sequential data. In HetEmotionNet, the LSTM is replaced with Gated Recurrent Units, while ATGRNet \cite{ATGRNet} uses a Temporal Convolutional Network (TCN). Some relatively simplified approaches, such as Siam-GCAN \cite{Siam-GCAN}, concatenate all the graph embeddings and feed them into a fully connected layer. MD-AGCN \cite{MD-AGCN} averages the graph embeddings of all time slices for further classification. \\

\noindent \textit{Remark.} Although the use of spatial-temporal graphs is currently limited, we argue this approach holds significant potential in this field. Emotions are dynamic processes that evolve with the development, intensification, and subsidence of internal and external stimuli \cite{houben2015relation, kragel2016decoding}. Therefore, this method aligns more closely with the neurophysiological foundations of emotions. However, current applications of spatial-temporal graphs have not deeply explored the construction of dynamic brain region relationships under emotional states. Existing methods typically treat the graph embeddings of EEG time slices as wholes to learn temporal dependencies between slices, thus limiting spatial dependencies to within individual slices. In other words, current research mainly focuses on the temporal variations of spatial relationships within isolated segments and is unable to construct connections between different nodes across slices.

\section{Future Directions}
In this section, we discuss some future research directions and possible approaches for graph application in EEG-based emotion recognition in addition to the challenges or limitations mentioned in the remarks in the previous sections.

\textbf{Temporal Graph:} This method aims to address the issue of incomplete temporal dependency present in current spatio-temporal graph methods, as discussed in Section~\ref{sec:stg}. This incomplete dependency refers to the lack of correlation between different channels across time slices. The spatio-temporal dependencies learned by existing methods are limited to the same electrode across different time slices, such as $v_i^{t-1}$ and $v_i^{t}$, while ignoring the relationships between different electrodes across time slices, such as $v_i^{t-1}$ and $v_j^{t}$. These inter-electrode relationships across time slices correspond to the delayed responses of brain regions in emotional states, indicating that interactions between brain regions exhibit asynchrony. The persistence of emotions involves asynchronous activities of brain regions, representing delayed responses of other regions to the current region's activity \cite{lewis2005bridging}.
Therefore, a potential direction is to establish a temporal graph between adjacent time slices, where edges exist between nodes across time slices rather than within the same time slice. In summary, computing a temporal graph for every pair of adjacent time slices can model the more intricate temporal dependencies of brain regions in emotional states. 

\textbf{Dependency-Level Interpretability:} Currently, interpretability in graph methods for this field has garnered little attention. However, given the physiological basis of emotion-related EEG, interpretability is a crucial direction for the future. We argue that the necessary interpretability methods in this field should focus on explaining the graph structure, specifically identifying which edges, rather than just nodes, are most important for emotion labels. Most interpretability methods are feature-level, such as \cite{chen2018learning,schlichtkrull2020interpreting} that determine which channels or time slices are most important for emotion prediction.
Due to the direct relationship between emotions and brain region dependencies, a promising approach is to highlight the significant edges, as illustrated in XGNN \cite{XGNN}, thereby reflecting crucial brain region relationships in emotion-related EEG. Another potential direction is using this method to learn interpretable sparse graphs. 
Similar to XGNN, by predicting how adding edges to the current graph affects the final label and setting an upper limit on the number of edges, it is possible to identify the most impactful edges for the final emotion label. These edges represent the sparse cortical activity between brain regions during emotional activities.

\textbf{Heterogeneous Graph for Mixed Emotion:} Although heterogeneous graphs have seen limited application in this field, we argue they represent a future mainstream trend, particularly as a promising approach for mixed emotion recognition. It is increasingly recognized that emotions are often mixed, meaning that subjects frequently experience two or more emotions simultaneously \cite{larsen2011further, cohen2009understanding,williams2002can}.
The potential of heterogeneous graphs lies in their ability to reveal the blurred boundaries between mixed emotional states more effectively than brain-centric emotional interaction models alone. By integrating multimodal physiological data, heterogeneous graphs build upon EEG-based brain emotion interaction models to construct comprehensive representations of emotional interactions within brain-based physiological systems. These interactions are strongly associated with emotion; for example, there is a correlation between prefrontal cortex oxygenation and decreased facial skin blood flow during positive emotions, whereas this correlation is less pronounced during negative emotions \cite{matsukawa2018deactivation}.
In summary, mixed emotion recognition is a crucial future research direction in this field, and the potential of heterogeneous graphs to elucidate the boundaries between mixed emotions makes it a method deserving of greater attention.

\section{Conclusion}
The rapid development of graph-based methods has revolutionized EEG-based emotion recognition. In this survey, we provide a comprehensive and updated review of graph techniques specifically designed for this field. We propose a novel taxonomy based on key components such as feature selection, edge computation, node aggregation, and graph manipulation. Our survey facilitates understanding the underlying mechanisms of applying graph methods to emotion recognition. Furthermore, we believe that consolidating the latest advancements and exploring future directions will inspire more innovative works within EEG-based emotion recognition.

\bibliographystyle{ACM-Reference-Format}
\bibliography{sample-base}

\appendix
\section{Appendix}
In this section, we provide a detailed categorized table as shown in Table~\ref{tab:baseline}.
\begin{table*}[ht]
\caption{Summary of graph-related methods for EEG-based emotion recognition}
\label{tab:baseline}
\resizebox{0.95\textwidth}{!}{
\begin{tabular}{c|cc|cccc|cc|ccc|c}
\hline  \hline 
\multirow{3}{*}{\textbf{Baseline}} &
  \multicolumn{2}{c|}{\textbf{Featrue Selection}} &
  \multicolumn{4}{c|}{\textbf{Edge Computation}} &
  \multicolumn{2}{c|}{\textbf{Node Aggregation}} &
  \multicolumn{3}{c|}{\textbf{Graph Manipulation}} &
  \multirow{3}{*}{\textbf{Year}} \\ \cline{2-12}
 &
  \multicolumn{1}{c|}{\multirow{2}{*}{\textbf{\begin{tabular}[c]{@{}c@{}}Temporal\\ Nodes\end{tabular}}}} &
  \multirow{2}{*}{\textbf{\begin{tabular}[c]{@{}c@{}}Frequency\\ Nodes\end{tabular}}} &
  \multicolumn{2}{c|}{\textbf{Model-independent Edge}} &
  \multicolumn{2}{c|}{\textbf{Model-dependent Edge}} &
  \multicolumn{1}{c|}{\multirow{2}{*}{\textbf{\begin{tabular}[c]{@{}c@{}}Spectral-\\ based\end{tabular}}}} &
  \multirow{2}{*}{\textbf{\begin{tabular}[c]{@{}c@{}}Spatial-\\ based\end{tabular}}} &
  \multicolumn{1}{c|}{\multirow{2}{*}{\textbf{\begin{tabular}[c]{@{}c@{}}Hierachical\\ Graph\end{tabular}}}} &
  \multicolumn{1}{c|}{\multirow{2}{*}{\textbf{\begin{tabular}[c]{@{}c@{}}Multi-\\ Graph\end{tabular}}}} &
  \multirow{2}{*}{\textbf{\begin{tabular}[c]{@{}c@{}}Spatial-Temporal\\ Graph\end{tabular}}} &
   \\ \cline{4-7}
 &
  \multicolumn{1}{c|}{} &
   &
  \multicolumn{1}{c|}{\textbf{Prior}} &
  \multicolumn{1}{c|}{\textbf{Signal Correlation}} &
  \multicolumn{1}{c|}{\textbf{Weighted}} &
  \textbf{Subspace} &
  \multicolumn{1}{c|}{} &
   &
  \multicolumn{1}{c|}{} &
  \multicolumn{1}{c|}{} &
   &
   \\ \hline
DGCNN \cite{DGCNN} &
   &
  DE,PSD,DASM,RASM,DCAU &
   &
  Gaussian Kernel &
   &
   &
  Polynomial &
   &
   &
   &
   &
  2018 \\
GCB-Net \cite{GCBNet} &
   &
  DE,PSD,DASM,RASM &
   &
  Gaussian Kernel &
   &
   &
  Polynomial &
   &
   &
   &
   &
  2019 \\
IAG \cite{IAG} &
   &
  DE &
   &
   &
   &
  Attention &
  Polynomial &
   &
  Brain Region Graph &
   &
   &
  2020 \\
RGNN \cite{RGNN} &
   &
  DE &
  Physical distance &
   &
   &
   &
  Polynomial &
   &
   &
   &
   &
  2020 \\
V-IAG \cite{V-IAG} &
   &
  DE,PSD,DASM,RASM &
   &
   &
   &
  Attention &
  Polynomial &
   &
  Brain Region Graph &
   &
   &
  2021 \\
SparseDGCNN \cite{Sparse-DGCNN} &
  ASM &
  DE,PSD,DASM,RASM,DCAU &
   &
  Gaussian Kernel &
   &
   &
  Polynomial &
   &
  Sparse Graph &
   &
   &
  2021 \\
HetEmotionNet \cite{HetEmotionNet} &
  Raw &
   &
   &
  Mutual Information &
   &
   &
  Linear &
   &
   &
  Temporal\&Frequency &
   &
  2021 \\
ASTG-LSTM \cite{ASTG-LSTM} &
   &
  DE,PSD &
   &
   &
   &
  Attention &
  Polynomial &
   &
   &
   &
  LSTM &
  2021 \\
MD-AGCN \cite{MD-AGCN} &
   &
  DE &
   &
  PCC &
   &
   &
  Linear &
   &
   &
  Temporal\&Frequency &
  Average &
  2021 \\
GECNN \cite{GECNN} &
   &
  DE,PSD,HHS &
   &
   &
   &
  Parameter Matrix &
  Polynomial &
   &
  Brain Region Graph &
   &
   &
  2021 \\
SOGNN \cite{SOGNN} &
  ASM &
  DE,PSD,DASM,RASM,DCAU &
   &
   &
   &
  Attention &
  Linear &
   &
  Sparse Graph &
   &
   &
  \multicolumn{1}{l}{2021} \\
GFIL \cite{GFIL} &
   &
  DE,PSD,DASM,RASM,DCAU &
   &
   &
   &
  Euclidean Distance &
  Linear &
   &
   &
   &
   &
  2021 \\
SWSC \cite{SWSC} &
   &
  DE,PSD,DASM,RASM,DCAU &
   &
   &
   &
  Euclidean Distance &
  Linear &
   &
   &
   &
   &
  2021 \\
Siam-GCAN \cite{Siam-GCAN} &
   &
  DE &
   &
   &
  Manhattan distance &
   &
  Polynomial &
   &
   &
   &
  Linear &
  2022 \\
Residual GCB-Net \cite{Residual-GCB-Net} &
   &
  DE,PSD,DASM,RASM &
   &
  Gaussian Kernel &
   &
   &
  Polynomial &
   &
   &
   &
   &
  2022 \\
SCC-MPGCN \cite{SCC-MPGCN} &
  Raw &
   &
   &
  PLV &
   &
   &
  Polynomial &
   &
  Brain Region Graph &
   &
   &
  2022 \\
OGSSL \cite{OGSSL} &
   &
  DE &
   &
  Euclidean Distance &
   &
   &
  Polynomial &
   &
   &
   &
   &
  2022 \\
MSTGNN \cite{MSTGNN} &
   &
  DE &
   &
  PLI &
   &
   &
   &
  MST &
   &
   &
   &
  2022 \\
JAGP \cite{JAGP} &
   &
  DE &
   &
   &
  Manhattan distance &
   &
  Polynomial &
   &
   &
   &
   &
  2022 \\
HD-GCN \cite{HD-GCN} &
   &
  DE &
  Brain Region &
   &
  Gaussian Kernel &
   &
  Linear &
   &
   &
  Local\&Global &
   &
  2022 \\
GMSS \cite{GMSS} &
   &
  DE &
  Brain Region &
   &
   &
   &
  Polynomial &
   &
  Brain Region Graph &
   &
   &
  2022 \\
TARDGCN \cite{TARDGCN} &
   &
  DE,Welch &
   &
  Cosine Similarity &
   &
   &
  Polynomial &
   &
   &
   &
   &
  2022 \\
AHGCN \cite{AHGCN} &
   &
  DE &
   &
   &
   &
  Parameter Matrix &
  Linear &
   &
  Sparse Graph &
   &
   &
  2022 \\
BLB-DGCNN \cite{BLB-DGCNN} &
  Raw &
   &
   &
  Gaussian Kernel &
   &
   &
  Polynomial &
   &
   &
   &
   &
  2022 \\
MD-GCN \cite{MD-GCN} &
   &
  DE &
  Pysical Distance &
   &
   &
   &
  Linear &
   &
   &
   &
   &
  2022 \\
DBGC-ATFFNetAFTL \cite{DBGC-ATFFNetAFTL} &
   &
  DE,PSD &
   &
   &
   &
  Parameter Matrix &
  Polynomial &
   &
   &
  Temporal\&Frequency &
   &
  2022 \\
CGCNN \cite{CGCNN} &
   &
  DE,PSD,DASM &
   &
  Logarithm &
   &
   &
  Polynomial &
   &
   &
   &
   &
  2022 \\
MDTDDL \cite{MDTDDL} &
   &
  DE,PSD &
   &
  Euclidean Distance &
   &
   &
  Polynomial &
   &
   &
   &
   &
  2022 \\
OMHGL \cite{OMHGL} &
   &
  DE,PSD &
   &
  Cosine Similarity &
   &
   &
  Polynomial &
   &
   &
   &
   &
  2023 \\
LGGNet \cite{LGGNet} &
  Raw &
   &
  Brain Region,Hemisphere &
   &
  Dot Product &
   &
  Linear &
   &
  Brain Region Graph &
   &
   &
  2023 \\
G2G \cite{G2G} &
   &
  DE,PSD &
  Node Connection &
   &
   &
   &
   &
  Attention &
   &
   &
   &
  2023 \\
STFCGAT \cite{STFCGAT} &
   &
  DE &
   &
  PLI &
   &
   &
   &
  Attention &
  Brain Region Graph &
   &
   &
  2023 \\
MESNP \cite{MESNP} &
   &
  PSD &
   &
  PLI &
   &
   &
  Polynomial &
   &
   &
   &
   &
  2023 \\
ST-GCLSTM \cite{ST-GCLSTM} &
   &
  DE &
   &
  PCC &
   &
   &
  Linear &
   &
   &
   &
  LSTM &
  2023 \\
DAGAM \cite{DAGAM} &
   &
  DE &
  Pysical Distance &
   &
   &
   &
  Linear &
   &
  Sparse Graph &
   &
   &
  2023 \\
DGGN \cite{DGGN} &
   &
  DE &
   &
  Gaussian Kernel &
   &
   &
  Linear &
   &
   &
   &
   &
  2023 \\
MSFR-GCN \cite{MSFR-GCN} &
   &
  DE &
   &
   &
  Euclidean Distance &
   &
  Linear &
   &
   &
   &
   &
  2023 \\
ST-SCGNN \cite{ST-SCGNN} &
   &
  DE,STFT &
   &
  PLV &
   &
   &
  Polynomial &
   &
   &
   &
   &
  2023 \\
SGLNet \cite{SGLNet} &
  Raw &
   &
  Pysical Distance &
   &
   &
   &
  Linear &
   &
   &
   &
  SNN,LSTM &
  2023 \\
GJFusion \cite{GJFusion} &
   &
  DE &
   &
  Mutual Information &
   &
   &
  Linear &
   &
  Sparse Graph &
   &
   &
  2023 \\
AMGCT \cite{AMGCT} &
   &
  DE &
   &
  Dot Product &
   &
   &
  Linear &
   &
   &
   &
   &
  2023 \\
ASGC \cite{ASGC} &
   &
  DE &
   &
   &
   &
  Parameter Matrix &
  Linear &
   &
   &
   &
   &
  2023 \\
ST-SCGNN \cite{ST-SCGNN} &
   &
  DE &
   &
  PLV &
   &
   &
  Polynomial &
   &
   &
   &
   &
  2023 \\
PGCN \cite{PGCN} &
   &
  DE,STFT &
  Pysical Distance &
   &
   &
  Attention &
  Linear &
   &
  Brain Region Graph &
   &
   &
  2023 \\
HN-DGST \cite{HN-DGST} &
   &
  DE,PSD,DASM,RASM,DCAU &
   &
  PCC &
   &
  Attention &
  Linear &
   &
   &
   &
   &
  2023 \\
MRGCN \cite{MRGCN} &
   &
  DE &
   &
  Euclidean Distance &
   &
   &
  Linear &
   &
   &
   &
   &
  2023 \\
ResGAT \cite{ResGAT} &
  Raw &
   &
   &
   &
   &
  Attention &
   &
  Attention &
   &
   &
   &
  2023 \\
MTLFuseNet \cite{MTLFuseNet} &
   &
  DE &
   &
  Mutual Information &
   &
   &
  Linear &
   &
   &
   &
   &
  2023 \\
GCNs–FSMI \cite{GCNs–FSMI} &
  Raw &
   &
   &
  PCC &
   &
   &
  Linear &
   &
   &
   &
   &
  2023 \\
LAG \cite{LAG} &
   &
  DE &
   &
  Dot Product &
   &
   &
  Polynomial &
   &
  Brain Region Graph &
   &
   &
  2024 \\
BF-GCN \cite{BF-GCN} &
   &
  DE &
   &
  PLV &
   &
  Parameter Matrix &
  Polynomial &
   &
   &
   &
   &
  2024 \\
PGCN \cite{PGCN2} &
   &
  DE,STFT &
  Node Connection &
   &
   &
  Attention &
  Linear &
   &
   &
  Local\&Global &
   &
  2024 \\
Gusa \cite{Gusa} &
   &
  DE,PSD,DASM,RASM,DCAU &
   &
   &
   &
  Attention &
  Polynomial &
   &
   &
   &
   &
  2024 \\
GDDN \cite{GDDN} &
   &
  DE,STFT &
   &
   &
  Concatenation &
   &
  Linear &
   &
   &
   &
   &
  2024 \\
BGAGCN \cite{BGAGCN} &
  Raw &
  DE,PSD &
  Node Connection &
  PCC &
   &
   &
  Polynomial &
   &
   &
  Local\&Global &
   &
  2024 \\
DG-JCA \cite{DG-JCA} &
   &
  DE,PSD,DASM,RASM,DCAU &
   &
  Logarithm &
   &
   &
  Linear &
   &
   &
   &
   &
  2024 \\
CU-GCN \cite{CU-GCN} &
   &
  DE,PSD,DASM,RASM &
  Node Connection &
   &
   &
   &
  Polynomial &
   &
   &
   &
   &
  2024 \\
ATGRNet \cite{ATGRNet} &
   &
  DE,PSD &
   &
   &
   &
  Parameter Matrix &
  Polynomial &
   &
   &
   &
  TCN &
  2024 \\
MD$^2$GRL \cite{MD$^2$GRL} &
   &
  DE,PSD &
  Pysical Distance &
   &
   &
   &
  Linear &
   &
  Sparse Graph &
  Temporal\&Frequency &
   &
  2024 \\
CCSR-GCN \cite{CCSR-GCN} &
   &
  DE &
   &
   &
   &
  Parameter Matrix &
  Polynomial &
   &
  Sparse Graph &
   &
   &
  2024 \\
Grop \cite{Grop} &
   &
  DE &
   &
  Gaussian Kernel &
   &
   &
  Polynomial &
   &
   &
   &
   &
  2024 \\
DS-AGC \cite{DS-AGC} &
   &
  DE &
   &
   &
  Euclidean Distance &
   &
  Polynomial &
   &
   &
   &
   &
  2024 \\
OnMHF \cite{OnMHF} &
  Raw &
  PSD &
   &
  Cosine Similarity &
   &
   &
  Linear &
   &
   &
   &
   &
  2024 \\
CGRU-MDGN \cite{CGRU-MDGN} &
   &
  DE &
   &
   &
  Concatenation &
   &
  Linear &
   &
   &
   &
   &
  2024 \\
VSGT \cite{VSGT} &
  Raw &
   &
   &
   &
  Concatenation &
   &
  Linear &
   &
   &
   &
  Temporal Graph &
  2024 \\
SSPA-GCN \cite{SSPA-GCN} &
  Raw &
   &
  Node Connection &
   &
   &
   &
  Polynomial &
   &
   &
   &
   &
  2024 \\
VBH-GNN \cite{VBH-GNN} &
  Raw &
   &
   &
   &
  Concatenation &
   &
  Linear &
   &
   &
   &
   &
  2024 \\ \hline \hline
\end{tabular}}
\end{table*}

\end{document}